\documentclass{article}

\usepackage{arxiv}
\usepackage{tcolorbox}
\usepackage[utf8]{inputenc} 
\usepackage[T1]{fontenc}    
\usepackage{hyperref}       
\usepackage{url}            
\usepackage{booktabs}       
\usepackage{amsfonts}       
\usepackage{nicefrac}       
\usepackage{microtype}      
\usepackage{lipsum}		
\usepackage{graphicx}
\usepackage{natbib}
\usepackage{float}
\usepackage{doi}

\title{System for Systematic Literature Review Using Multiple AI agents: Concept and an Empirical Evaluation}

\author{ \href{https://orcid.org/0000-0000-0000-0000}{\includegraphics[scale=0.06]{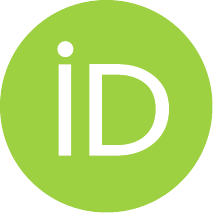}\hspace{1mm}Abdul Malik Sami}\\
        Tampere University\\
	\texttt{malik.sami@tuni.fi} \\
	\And
	\href{https://orcid.org/0000-0000-0000-0000}{\includegraphics[scale=0.06]{orcid.pdf}\hspace{1mm}Zeeshan Rasheed} \\
	Tampere University\\
	\texttt{zeeshan.rasheed@tuni.fi} \\
	\AND
	Kai-Kristian Kemell \\
	University of Helsinki \\
	\texttt{kai-kristian.kemell@helsinki.fi} \\
	\And
	Muhammad Waseem \\
	Jyväskylä University \\
	\texttt{muhammad.m.waseem@jyu.fi} \\
	\And
	Terhi Kilamo \\
	Tampere University \\
	\texttt{terhi.kilamo@tuni.fi} \\
        \And
	Mika Saari \\
	Tampere University \\
	\texttt{mika.saari@tuni.fi} \\
        \And
	Kari Systä \\
	Tampere University \\
	\texttt{kari.systa@tuni.fi} \\
        \And
	Anh Nguyen Duc \\
	University of South Eastern Norway \\
	\texttt{Anh.Nguyen.duc@usn.no} \\
        \And
	Pekka Abrahamsson \\
	Tampere University \\
	\texttt{pekka.abrahamsson@tuni.fi} \\
}

\renewcommand{\shorttitle}{\textit{arXiv} Template}

\begin{document}
\maketitle

\begin{abstract}

Systematic literature review (SLR) is foundational to evidence-based research, enabling scholars to identify, classify, and synthesize existing studies to address specific research questions. Conducting an SLR is, however, largely a manual process. In recent years, researchers have made significant progress in automating portions of the SLR pipeline to reduce the effort and time required for high-quality reviews; nevertheless, there remains a lack of AI-agent-based systems that automate the entire SLR workflow.
To this end, we introduce a novel multi-AI-agent system designed to fully automate SLRs. Leveraging large language models (LLMs), our system streamlines the review process to enhance efficiency and accuracy. Through a user-friendly interface, researchers specify a topic; the system then generates a search string to retrieve relevant academic papers. Next, an inclusion/exclusion filtering step is applied to titles relevant to the research area. The system subsequently summarizes paper abstracts and retains only those directly related to the field of study. In the final phase, it conducts a thorough analysis of the selected papers with respect to predefined research questions.
This paper presents the system, describes its operational framework, and demonstrates how it substantially reduces the time and effort traditionally required for SLRs while maintaining comprehensiveness and precision. The code for this project is available at: https://github.com/GPT-Laboratory/SLR-automation
.


\end{abstract}


\section{Introduction}
\label{Introduction}
The Systematic Literature Review (SLR) is a fundamental component of academic research, offering a comprehensive and unbiased overview of existing literature on a specific topic \cite{keele2007guidelines}. It involves a structured methodology for identifying, evaluating, and synthesizing all relevant research to address clearly defined research questions \cite{kitchenham2009systematic}. This process is critical for establishing the context and foundation of new research, identifying gaps in current knowledge, and informing future research directions \cite{van2021automation}. However, conducting an SLR is inherently time-consuming and labor-intensive. It requires meticulous planning, extensive searching, and rigorous screening of large volumes of literature. The complexity and scale of this task, especially in fields with vast and rapidly expanding bodies of work, can be daunting and resource-intensive. The challenge lies not only in the collection of relevant literature but also in the accurate synthesis and interpretation of the gathered data.

The emergence of Large Language Models (LLMs) in Artificial Intelligence (AI) presents new opportunities for automating and streamlining the SLR process \cite{rasheed2024codepori}, \cite{rasheed2023autonomous}. LLMs, trained on extensive datasets of text, are adept at understanding and generating human-like language \cite{carlini2021extracting}. They can process and analyze large volumes of text rapidly, offering insights and summaries that would take humans significantly longer to compile. Their ability to understand context and nuances in language makes them particularly useful for tasks like identifying relevant literature, extracting key information, and summarizing research findings \cite{hou2023large}. By automating the more tedious and repetitive aspects of the SLR process, LLMs can significantly reduce the time and effort required, allowing researchers to focus on the more nuanced aspects of their research \cite{rasheed2024can}.

In this context, our proposed system utilizes the capabilities of LLMs to automate the whole SLR process. We developed a LLM-based multi-agent system  that automates each step of the SLR, from the initial literature search to the final analysis. The system begins with a simple user input–researchers enter their topic into a designated text box. This input is then processed by the LLM, which generates a precise search string tailored to retrieve the most relevant academic papers. The system's next phase involves an intelligent filtering mechanism. It applies an inclusive and exclusive theory, screening titles and abstracts to retain only those studies that are directly relevant to the specified research area. 

The final stage of our system autonomously summarizes the abstracts of the selected papers, ensuring that only content pertinent to the research questions is retained. It introduces a level of precision and consistency in data analysis that is challenging to achieve manually. Finally, the proposed system conducts an in-depth analysis of the selected papers, aligning its findings directly with the research questions. This comprehensive approach ensures that the final output is not only a reflection of the vast array of literature available but also a focused and relevant resource tailored to the specific needs of the researcher. Our system, therefore, stands as a testament to the potential of integrating advanced AI technologies in academic research methodologies.






\section{Related Work}
\label{Related Work}

\cite{bartholomew2002james} conducted the first SLR to carried out systematic clinical trials to identify effective treatments for scurvy. His trials, which rigorously evaluated various potential remedies, notably highlighted the effectiveness of oranges and lemons as the most successful treatments \cite{bartholomew2002james}. In the domain of SE research, the SLR approach was introduced by \cite{kitchenham2004procedures}. This framework was instrumental in adapting the principles of systematic reviews, already prevalent in fields like healthcare and social sciences, to the specific challenges and needs of SE research. Following this development, SLRs have become an extensively used practice to support evidence-based material in SE. The success of SLRs in facilitating evidence-based studies has motivated other researchers to adopt this approach in their work \cite{kitchenham2009systematic}. However, Undertaking SLRs is often a challenging endeavor, encompassing various activities such as gathering, assessing, and recording evidence. These tasks within SLRs are typically done manually, without the aid of automation or decision support tools, making the process not only time-intensive but also susceptible to errors \cite{van2021automation}. Many researchers make progress to automate the process of SLR \cite{van2021automation}.

Current research efforts are primarily focused on refining the SLR process to optimize precision while ensuring high recall, addressing the precision shortcomings often found in existing methods \cite{o2015using}. Additionally, there's a significant push towards reducing human errors, especially since many steps in the review process are highly repetitive \cite{marshall2016tool}. In this context, the works of K.R. Felizardo and J.C. Maldonado are notable. They have explored the shift from traditional, repetitive, and error-prone SLR methods towards the application of visual text mining. This approach, as outlined in their articles \cite{felizardo2012visual}, \cite{felizardo2014visual}, \cite{felizardo2011analysing}, \cite{malheiros2007visual} leverages unsupervised learning to assist users in identifying relevant articles, though it does require users to have a background in machine learning and statistics.

\cite{olorisade2016critical} presented an innovative ML model designed to automate the primary study selection process in SLRs, potentially streamlining this critical step and significantly reducing the manual effort involved in sifting through vast quantities of academic literature. \cite{shakeel2018automated} provided valuable insights into potential threats that could arise when automating the SLR process. \cite{feng2017text} highlighted various text mining techniques currently employed in SLRs, a foundation upon which our tool builds. Significantly, \cite{paynter2016epc} presented a comprehensive report delineating the application of text mining (TM) techniques in automating various stages of the SLR process, including selection, extraction, and updates. This aligns closely with our tool's objectives. \cite{clark2020full} demonstrated the feasibility of completing an SLR in a markedly reduced time frame using multiple tools, a precedent for the efficiency our tool aims to achieve.

\cite{michelson2019significant} provided an economic analysis and time estimates for SLRs, underscoring the need for automated solutions – a call that our tool directly responds to. In a similar vein, \cite{beller2018making} not only listed tools useful for automating SLRs but also established eight guidelines that have informed our tool’s development.

\cite{jonnalagadda2015automating} detailed methods for data extraction from published reports, which has been instrumental in shaping our tool’s data handling capabilities. Moreover, \cite{marshall2019toward} and \cite{o2019question} have respectively listed useful tools for systematic reviews and articulated barriers to the adoption of such tools, providing a comprehensive understanding of the current landscape and user hesitance in this domain. Further contributions include \cite{o2015using} and \cite{o2015using}, who respectively conducted an SLR on text mining in the automation of SLRs and described the automation potential across different steps in the SLR process. These works have been pivotal in identifying areas where our tool can be most impactful. Additionally, \cite{jaspers2018machine} and \cite{thomas2011applications} have explored machine learning techniques and the application of TM techniques in automating the SLR process, which have been key influences in our tool's design. Lastly, the survey by \cite{van2019usage}, highlighting the limited use of SLR tools among researchers, emphasizes the need for more user-friendly and efficient solutions like the one our tool aims to provide.

Despite these advancements in automating the SLR process, there remains a notable gap in the complete automation of SLRs using LLMs. Addressing this gap, we have developed a novel approach: a multi-agent system based on LLMs. This innovative system is designed to fully automate the SLR process, utilizing the advanced capabilities of LLMs to efficiently manage and synthesize vast amounts of data, which is a significant step forward in the field of automated literature reviews.

\section{Research Method}
\label{Research Method}
This research aims to investigate how an LLM-based multi-agent system can be utilized to automate the entire process of SLRs. Below, we discuss how our LLM-based multi-agent system collaborates and performs such tasks. We have formulated a research questions (RQ):

\begin{tcolorbox}[colback=green!2!white,colframe=black!75!black]
\textit{\textbf{RQ.} How does a LLM-based multi-agent system transform traditional methodologies to automate the systematic literature review process in SE?}
\end{tcolorbox}
\textbf{Motivation}: The motivation for the research question arises from the need to enhance the efficiency and effectiveness of literature review processes in the rapidly evolving field of SE. Traditional methods of conducting literature reviews are often time-consuming and labor-intensive, potentially leading to delays in research progress and the dissemination of new knowledge. The integration of LLMs-based agents promises a paradigm shift, potentially automating and streamlining these processes. By exploring the transformation brought about by an LLM-based multi-agent system, this research seeks to reduce the time and effort required for comprehensive literature reviews and also to increase the accuracy and scope of these reviews. This could result in more timely and informed research outcomes in SE, a field where staying abreast of current trends, methodologies, and discoveries are crucial for technological advancement and innovation.


\subsection{LLM-based Agent Assisted Systematic Literature Review}
\label{LLM-Based Assisted Systematic Literature Review}

The section focuses on research methodology for developing an LLM-based multi-agent system. This system is specifically engineered to automate the entire process of conducting SLRs. The innovation lies in its ability to transform a given research topic into a comprehensive review through a series of automated, interconnected steps. Each step is managed by a specialized agent within the system, working collaboratively to ensure a seamless and efficient literature review process. In Figure \ref{fig: multi-agent system}, we illustrate how agents collaborate with each other to generate a response. Below, we also provide the detailed functionality of each agent within this multi-agent system.

\subsubsection{Planner agent}

The first agent is dedicated to generating research questions and purpose and search strings. Upon receiving a research objective from the end-user, For the search string we provide research questions and objective to the agent, it generates a search string. this agent employs advanced language understanding algorithms to interpret the topic's key elements. The system, designed for deep semantic understanding, analyzes the topic to extract key concepts, themes, and terminologies. It then utilizes its extensive training on diverse textual data to construct a precise and comprehensive search string. This string is formulated by combining relevant keywords, synonyms, and technical terms that capture the essence of the research question. Furthermore, the algorithm is adept at understanding context and varying semantic structures, enabling it to refine the search string to match specific research domains. The generated search string is crucial in accurately retrieving relevant literature from various academic databases. By ensuring that the initial search is both thorough and focused, the agent significantly enhances the efficiency and quality of the literature collection process. This sets a solid foundation for the subsequent stages of the SLR, where the depth and breadth of the collected literature play a crucial role.

\subsubsection{Literature identification agent}

Following the generation of the search string, research questions, and the purpose of each question, the next agent takes over the task of literature retrieval. This agent is responsible for using the search string to query academic databases and retrieving initial sets of papers that are potentially relevant to the research topic. It employs sophisticated filtering algorithms to manage the vast amount of available data, selecting papers by their title which are most closely align with the predefined parameters of the search string. This step is crucial in narrowing down the pool of literature to a manageable size for in-depth review.

\begin{figure*}[t]
    \centering
    \includegraphics[width=0.8\textwidth]{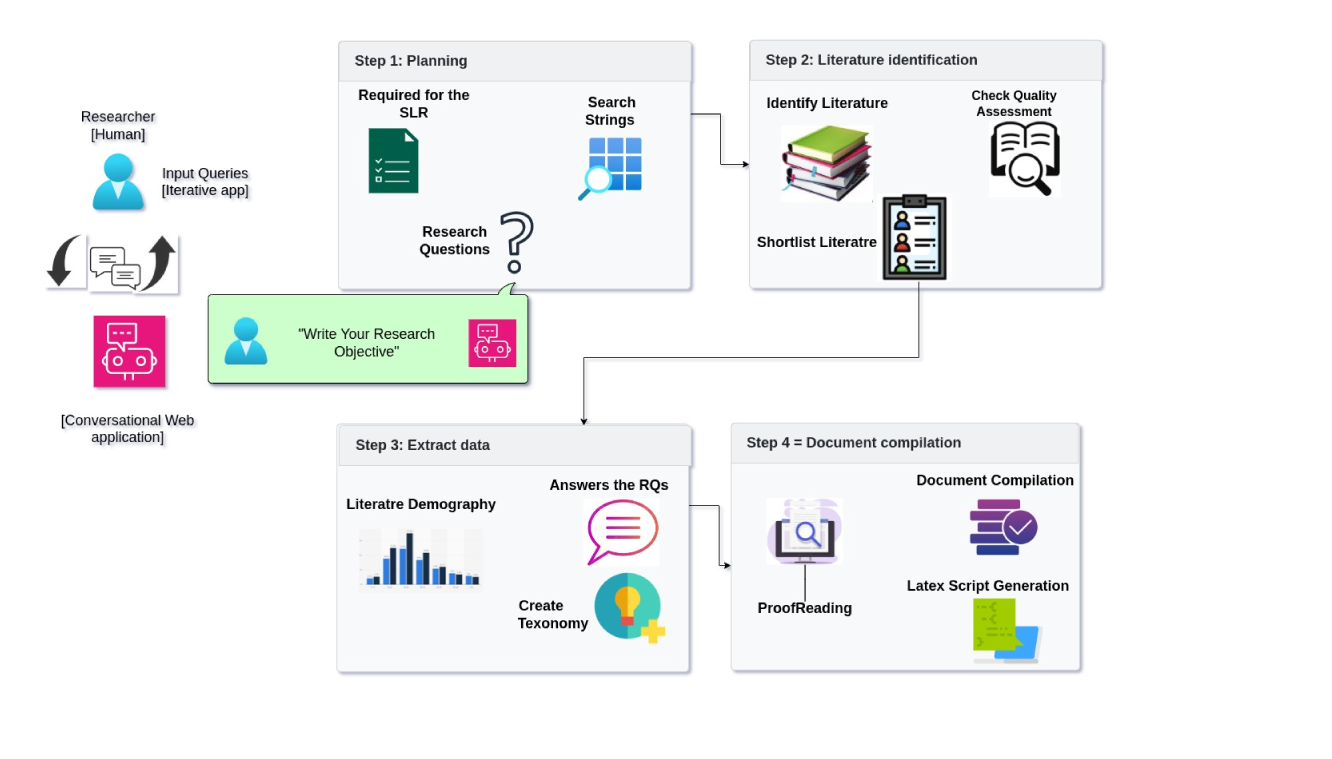}
    \caption{Process of Systematic Literature Review Automation Tool}
    \label{fig:result}
\end{figure*}

\subsubsection{Data extraction agent}

The third agent in our system is tasked with refining the literature using inclusion and exclusion criteria based on the research objectives. Initially, it employs our LLM algorithm to analyze the titles of retrieved papers, discerning their relevance to the research topic. This step involves text analysis, where the LLM algorithm identifies key terms and concepts that align with the research objectives. By applying these predefined rules, the agent effectively filters out irrelevant material, ensuring the literature review remains focused and pertinent to the research questions.

Following the title analysis, the agent proceeds to analyze the abstracts of selected papers. LLM algorithm conducts a more in-depth text analysis, evaluating the context, methodologies, and findings in the abstracts to assess their relevance. The final and most comprehensive step involves analyzing the full content of each paper. This thorough examination encompasses the whole paper, allowing the agent to evaluate how each paper’s content and findings relate to the specific research question. 
This agent extracts key information, Answers each question of the filtered papers, and shows its data in tabular form, a. It then synthesizes this information to provide a comprehensive overview of the current state of research on the topic. This synthesis is crucial in understanding the broader context and implications of the findings within the selected literature.

\subsubsection{Data compilation agent}

The final agent in our multi-agent system is responsible for analyzing the synthesized data in relation to the research questions and objectives. It assesses trends, identifies gaps in the literature, and draws conclusions based on the aggregated information. This agent also prepares a report that summarizes the findings of the literature review, providing a clear and concise overview of the research landscape for the given topic.

Each agent in the LLM-based multi-agent system plays a vital role in automating the systematic literature review process. From generating search strings to reporting findings, the agents work in a coordinated manner to ensure a thorough, efficient, and accurate review. This methodology represents a significant advancement in the way SLRs are conducted, offering a more streamlined and effective approach to academic research.

\section{Results}
\label{Result}
This section presents the results obtained from implementing an AI-agent-based system aimed at automating the SLR process in SE. The findings are provided below in detail.

\subsection{LLM Based Multi-Agent System (RQ)}
\begin{figure}[H]
    \centering
    \includegraphics[width=0.9\columnwidth]{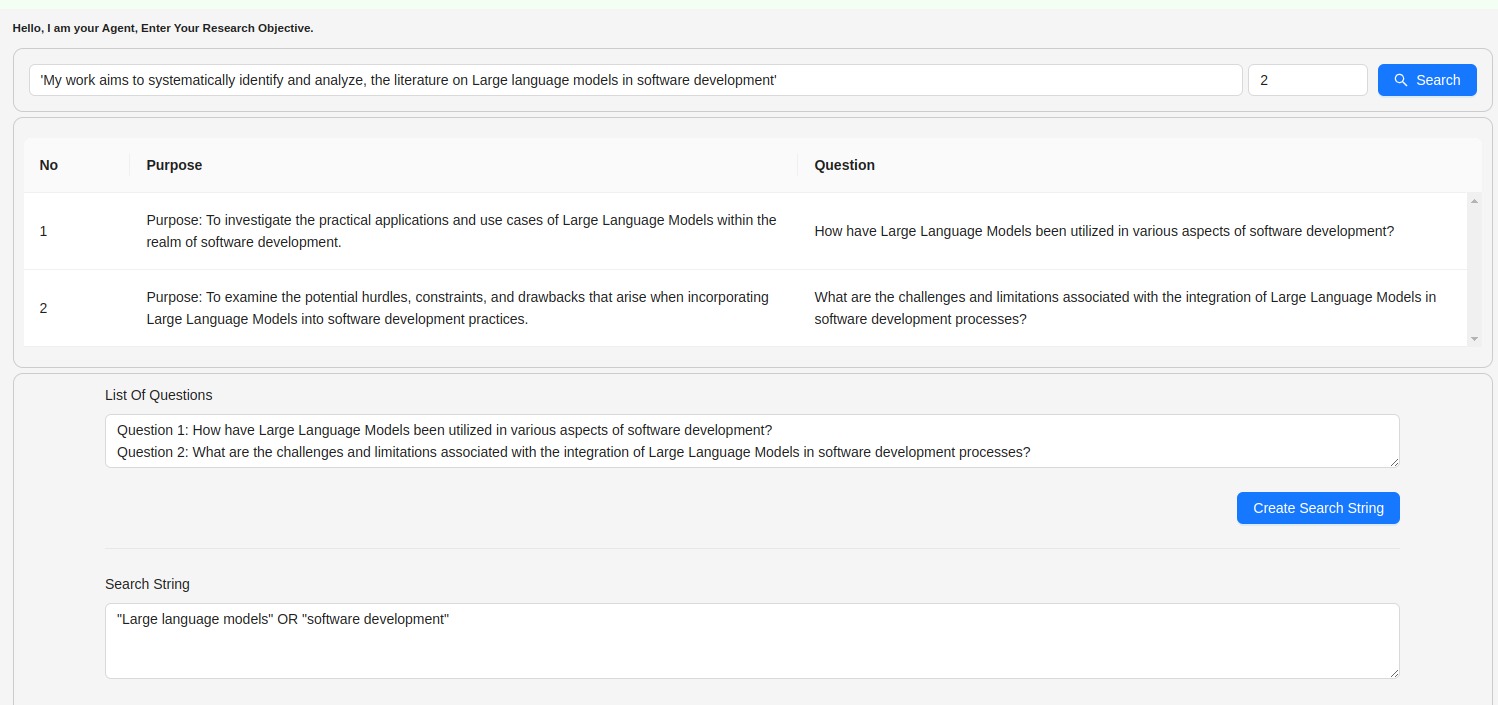}
    \caption{Results of Search String and RQs}
    \label{fig:search string}
\end{figure}

Our research introduces an LLM-based multi-agent system that redefines the conventional approach to SLRs in SE. The multi-agent system developed for automating the SLR process has demonstrated its efficacy through a structured and sequential workflow. The process begins with the input of a research topic, as depicted in Figure \ref{fig:search string}.

\begin{figure}[H]
   \centering
    \includegraphics[width=0.9\columnwidth]{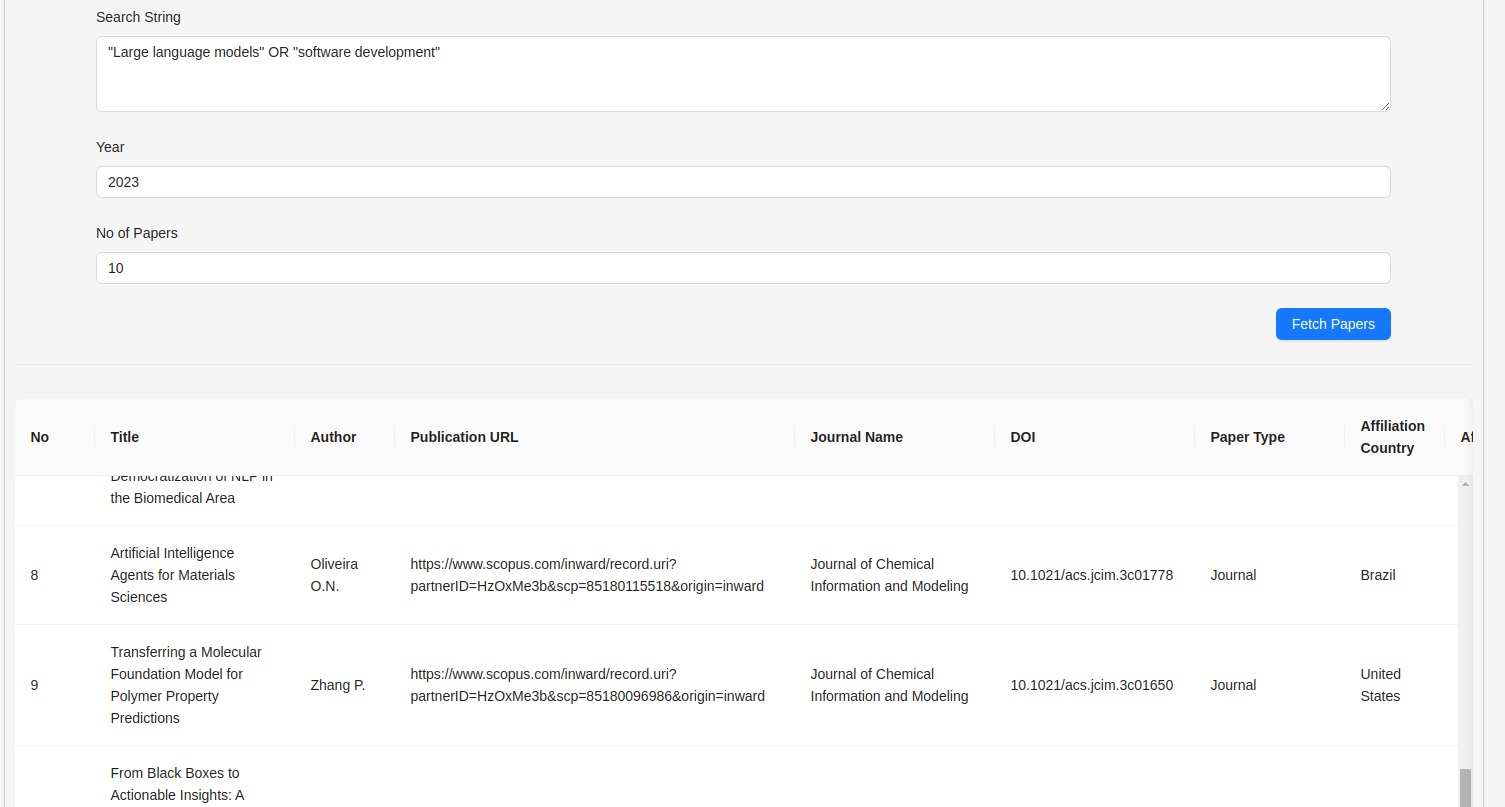}
    \caption{Results of Paper Identification}
    \label{fig:paper identification}
\end{figure}

Upon receiving the topic, the system systematically generates a pertinent set of research questions. As illustrated in Figure \ref{fig:search string}, it formulated questions like 'How have large language models been utilized in various aspects of the software development process?' and 'What challenges and limitations exist in the adoption and implementation of large language models in software development?' These questions play a crucial role in guiding the literature search and analysis process. Subsequent to the research question formulation, the system proceeds to generate a search string. In this case, the search string ``large language models OR software development'' was created, coupled with a specified year to narrow down the search scope, thereby enhancing the relevance and precision of the search results.

\begin{figure}[H]
   \centering
    \includegraphics[width=0.9\columnwidth]{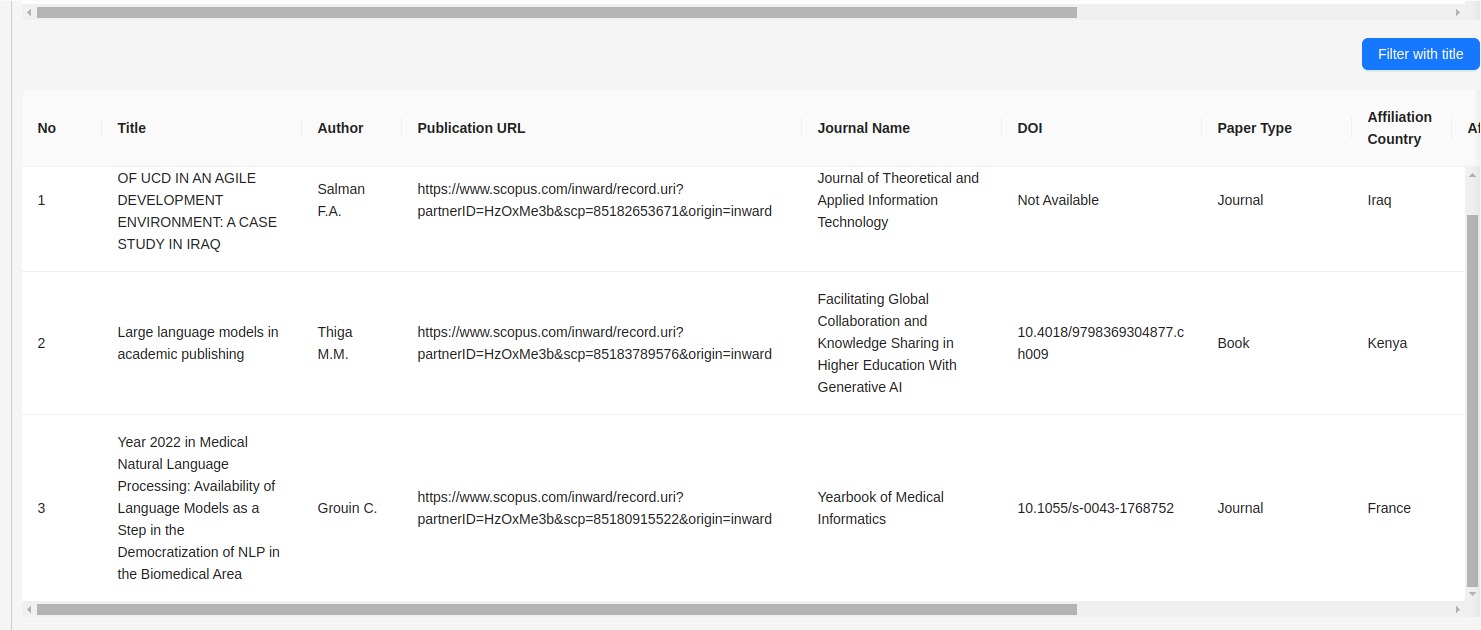}
    \caption{Paper Filtering}
    \label{fig:paper filtering}
\end{figure}

The subsequent phase involves retrieving papers that match the generated search string. This feature of the system is specifically designed to fetch research papers from various databases, as demonstrated by the list of papers shown in Figure \ref{fig:paper identification}. For this demonstration, we focused solely on papers published in the year 2023, setting the system to retrieve only 10 papers from that year, all relevant to this field. The tool efficiently compiles relevant information such as the title, author, publication URL, journal name, DOI, paper type, affiliation country, and the affiliation institution. Moreover, the system is equipped with the capability to apply inclusive and exclusive criteria based on titles, which further refines the search results to ensure only the most pertinent literature is considered for review. As shown in Figure \ref{fig:paper filtering}, only three papers were selected for an in-depth analysis.

Finally, the system extracts data based on the formulated RQs. This advanced feature is exemplified in the demo where detailed answers are provided for the previously generated RQs. For instance, the answer to the first RQs discusses the varied applications of large language models in the software development lifecycle and highlights specific instances of their use, like the inference from the paper ``InferLink End-to-End Program Repair with Large Language Models.''

\begin{figure}[H]
   \centering
    \includegraphics[width=0.9\columnwidth]{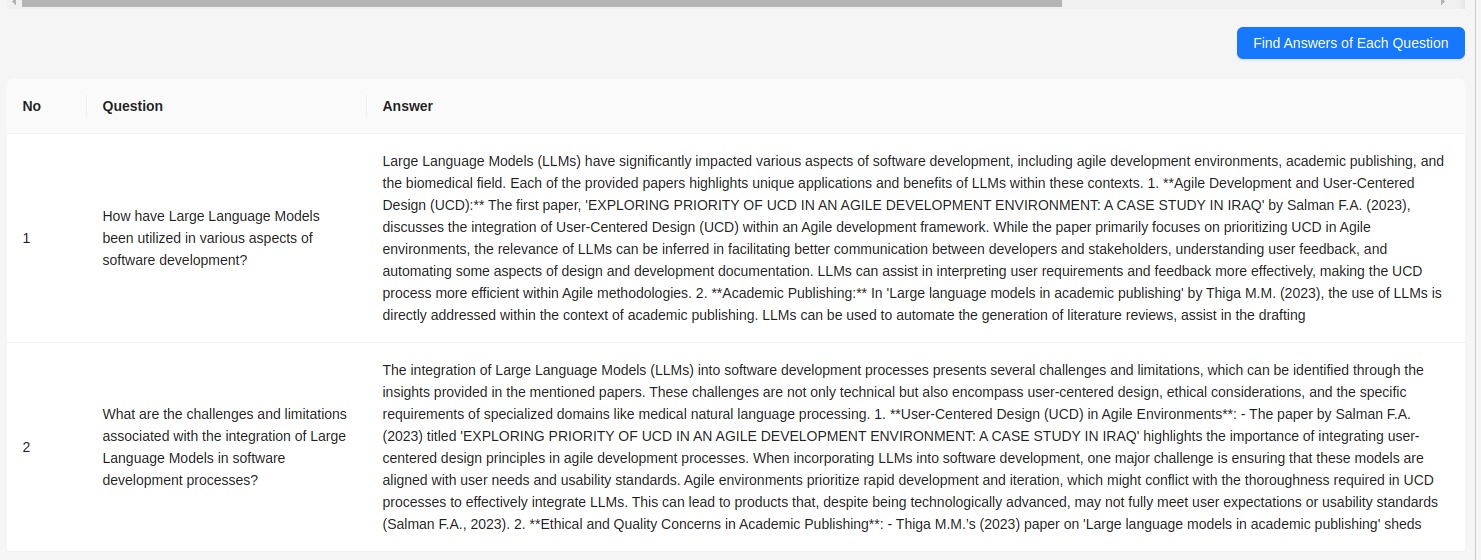}
    \caption{Data Extraction}
    \label{fig:data extraction}
\end{figure}

In conclusion, the proposed system has showcased its ability to streamline the laborious process of literature review, from defining the scope of the research to extracting and synthesizing data pertinent to the research questions. The demonstration affirms the system's potential in significantly reducing the time and effort conventionally required for conducting systematic literature reviews.

\begin{figure}[H]
   \centering
    \includegraphics[width=0.9\columnwidth]{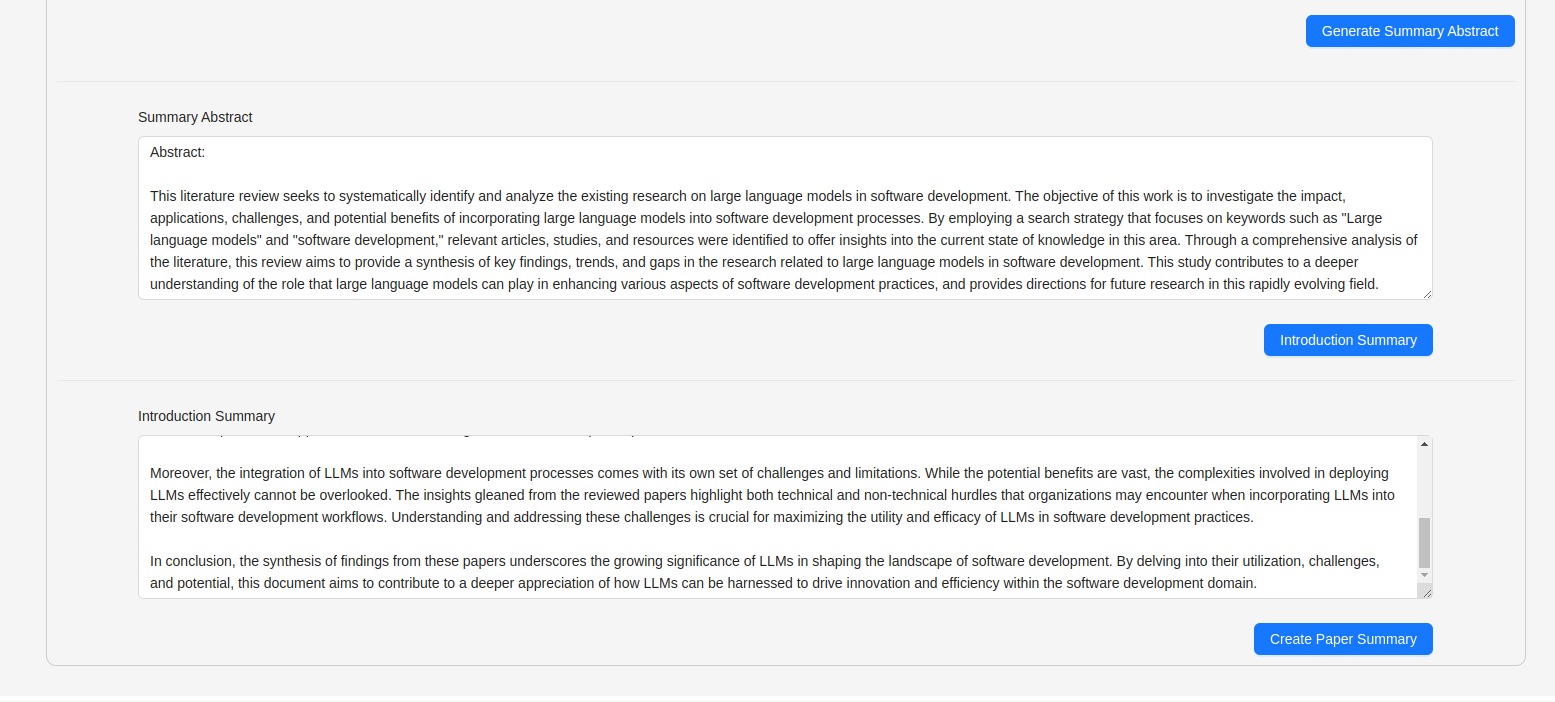}
    \caption{Summary of Abstract and Introduction}
    \label{fig:sumary}
\end{figure}

\section{Discussions}
\label{Discussion}

The results derived from the implementation of our multi-AI agent system for SLR have been both encouraging and insightful. The system successfully automated key components of the SLR process, including the generation of search strings, the selection and filtering of relevant literature, and the summarization of key findings. This automation significantly reduced the time and effort typically required in conducting SLRs, while maintaining, and in some aspects enhancing, the accuracy and comprehensiveness of the review. The system’s ability to process and analyze large volumes of text rapidly, and its precision in identifying relevant studies, demonstrated the substantial potential of integrating LLMs in academic research.

The implications of these results are far-reaching. Firstly, the system presents a valuable tool for researchers across various fields, reducing the barriers to conducting comprehensive literature reviews. This efficiency can accelerate the pace of research and discovery, enabling scholars to focus on more complex and creative aspects of their work. Additionally, the system’s standardization of the SLR process can potentially lead to more consistent and replicable research outcomes, a cornerstone in scientific research. The reduction in manual effort also opens up opportunities for researchers with limited resources or those facing constraints such as time pressures, broadening the scope of who can conduct thorough literature reviews.

In addition, the long-term impact of our work on researchers and the broader academic community is expected to be substantial. Our system represents a paradigm shift in how SLRs are conducted, offering a tool that is not only efficient but also adaptable to the evolving needs of researchers. One significant future impact is the democratization of research. By simplifying the SLR process, our tool makes high-quality literature reviews accessible to a wider range of researchers, including those from institutions with fewer resources or those new to the field. This accessibility could lead to a more diverse range of voices and perspectives in academic research, enriching the field as a whole.

Furthermore, the system's efficiency in handling large volumes of data makes it an invaluable asset in fields where literature is vast and rapidly growing, such as biomedical research, environmental studies, and technology. Researchers in these fields can stay abreast of the latest developments more effectively, ensuring their work is informed by the most current and comprehensive data available.

In the domain of interdisciplinary research, our system can facilitate the synthesis of information across different fields, potentially leading to novel insights and innovations. By efficiently collating and analyzing diverse sets of literature, the tool can help uncover connections between disciplines that might otherwise be overlooked.

The long-term adaptation of our system based on user feedback and technological advancements will also ensure its ongoing relevance. Continuous updates will allow the system to incorporate the latest AI advancements, further enhancing its capabilities and ensuring it remains a cutting-edge tool for SLR. Moreover, the system's potential for customization will allow it to cater to the specific needs of different research domains. This customized approach means that the tool can be fine-tuned to deliver more targeted and relevant results, depending on the specific requirements of the research question or field.

\section{Limitation}
This study, while contributing valuable insights into the SE field. However, there is several limitations that necessitate attention in future iterations of the research. Primarily, the initial search strategy employed for identifying relevant literature was sub optimal. The absence of a comprehensive use of Boolean operators, notably the lack of "AND" in the search strings, potentially compromised the specificity and thoroughness of the literature search, leading to an incomplete representation of the available evidence. This issue underscores the need for a more rigorously defined search strategy to enhance the precision and relevance of retrieved documents.

Furthermore, the methodology exhibited a significant gap in its approach to literature selection, characterized by an absence of clearly defined criteria for primary and secondary exclusion. This oversight likely resulted in a less rigorous filtering process, diminishing the study's ability to exclude irrelevant or low-quality studies systematically. Implementing explicit inclusion and exclusion criteria will be crucial for improving the reliability and validity of the literature review in subsequent versions of the paper.

Another critical limitation observed was in the data extraction phase. Although data were extracted based on predefined research questions, the reliability of the extracted information is questionable due to the lack of a robust analytical algorithm. The current methodology does not adequately ensure the accuracy and relevance of the extracted data, which is a cornerstone for drawing reliable conclusions. Future iterations of this research will benefit substantially from the integration of advanced analytical algorithms capable of more advanced data analysis. Such algorithms should not only extract data more efficiently but also evaluate the quality and applicability of the information in relation to the research objectives.

Addressing these limitations is essential for advancing the research's contribution to the field. Enhancements in search strategy, literature screening, and data analysis will not only refine the methodological approach but also improve the study's overall credibility and impact. Future work will focus on fixing these issues to establish a more reliable and comprehensive research framework.

\section{Future Work}
Addressing the identified limitations presents a pathway for enhancing the comprehensiveness of our research in future iterations. The upcoming version of this paper will aim to implement several key improvements.

\textbf{Refinement of search strategy:} To overcome the limitations posed by an inadequate search string, future work will involve the development of a more sophisticated search strategy. This will include the comprehensive use of Boolean operators, particularly the incorporation of "AND" to ensure the specificity and thoroughness of the literature search. A systematic approach to defining search strings will be adopted to enhance the precision and relevance of retrieved documents.

\textbf{Implementation of explicit exclusion and inclusion criteria:} Recognizing the absence of clearly defined criteria for primary and secondary literature exclusion as a significant gap, future efforts will focus on establishing explicit inclusion and exclusion criteria. This refinement will facilitate a more rigorous and systematic screening process, thereby improving the study's ability to exclude irrelevant or low-quality studies systematically and ensuring a more reliable and valid literature review.

\textbf{Advancement of data extraction methods:} The preliminary phase highlighted the need for a more reliable data extraction mechanism. To address this, future work will incorporate advanced analytical algorithms designed to ensure the accuracy and relevance of the extracted data. These algorithms will not only facilitate more efficient data extraction but will also provide a means to critically evaluate the quality and applicability of the information in relation to the research objectives. The integration of machine learning and natural language processing techniques will be explored to automate and enhance the data extraction and analysis process.

\textbf{Enhancement of analytical framework:} Acknowledging the limitations in the initial data analysis, future research will aim to develop and implement a more robust analytical framework. This framework will be designed to analyze the extracted data comprehensively, incorporating both qualitative and quantitative methodologies as appropriate. Emphasis will be placed on ensuring the reliability and validity of the findings through rigorous statistical testing and sensitivity analyses.

\textbf{Broadening of literature scope:} To counteract any potential biases or gaps in the literature review caused by the initial search limitations, future research will broaden its scope to include a wider range of databases and grey literature. This expansion will ensure a more comprehensive coverage of the subject matter, encompassing diverse perspectives and emerging research trends.

\textbf{Stakeholder engagement:} Recognizing the value of stakeholder insights in refining research methodologies, future iterations will involve engaging with domain experts, researchers, and practitioners. This engagement will provide critical feedback on the research design, methodologies, and findings, contributing to a more nuanced and impactful research outcome.

By systematically addressing these limitations, future work will significantly enhance the study's contribution to the field, providing a more robust, comprehensive, and reliable foundation for understanding the research topic. These improvements will not only address the current study's shortcomings but also set a precedent for methodological rigor in similar research endeavors.
\section{Conclusions}
\label{Conclusion}
The development and implementation of our multi-AI agent system represent a significant advancement in the field of SLR. By integrating the capabilities of LLMs, this research demonstrates a novel approach to automating and optimizing the SLR process. Our system addresses the primary challenges associated with traditional SLR methods: the time-consuming nature of the process and the potential for human error or bias in literature selection and analysis. By automating the initial search, screening, summarization, and analysis phases, the system significantly reduces the manual effort and time required, while also enhancing the accuracy and consistency of the results.

The use of a simple user interface for topic input and subsequent generation of tailored search strings illustrates the system's user-friendly approach, making complex SLR processes accessible to a broader range of researchers. The inclusive and exclusive filtering mechanism ensures that the literature review remains focused and relevant, directly aligning with the specified research questions. The autonomous summarization of abstracts and the final analytical phase underscore the system's ability to extract vast amounts of data into clear, relevant information, a task that would be challenging without the aid of advanced AI.

This research contributes to the growing field of AI application in academic research, showcasing how LLMs can be effectively employed to enhance research methodologies. While the system significantly improves efficiency and accuracy, it is important to acknowledge the role of human oversight in guiding and interpreting the results, ensuring that the final output maintains the depth required in scholarly research.

\bibliographystyle{unsrtnat}
\bibliography{references}  






\end{document}